\documentclass[conference]{IEEEtran}
\IEEEoverridecommandlockouts
\usepackage{cite}
\usepackage{amsmath,amssymb,amsfonts}
\usepackage{algorithmic}
\usepackage{graphicx}
\usepackage{textcomp}
\usepackage{xcolor}
\usepackage{textcomp}
\usepackage{multirow}
\usepackage{subfigure}
\usepackage{xcolor}
\usepackage{multirow}
\usepackage{amsmath,amssymb,amsfonts}
\usepackage{xcolor,color,colortbl}
\usepackage{amsmath,amssymb,amsfonts}
\usepackage{algorithm}
\usepackage{enumitem}
\usepackage{hyperref}

\newcommand\0{\kern-1.2pt\vec{\kern1.2pt 0}}
\newcommand{\bigO}{\mathcal{O}}
\usepackage{enumitem}
\usepackage{mathtools}

\usepackage{cite}
\usepackage{balance}
\def\BibTeX{{\rm B\kern-.05em{\sc i\kern-.025em b}\kern-.08em
    T\kern-.1667em\lower.7ex\hbox{E}\kern-.125emX}}

\begin{document}

\title{Joint Estimation of Multiple RF Impairments Using Deep Multi-Task Learning}

\author{\IEEEauthorblockN{Mehmet Ali Ayg\"{u}l\IEEEauthorrefmark{1}\IEEEauthorrefmark{2}, 
Ebubekir Memi\c{s}o\u{g}lu\IEEEauthorrefmark{3},~H\"{u}seyin~Arslan\IEEEauthorrefmark{3}\IEEEauthorrefmark{4}}
\IEEEauthorblockA{\IEEEauthorrefmark{1}Department of Electronics and Communications Engineering, Istanbul Technical University, Istanbul, 34467 Turkey}
\IEEEauthorblockA{\IEEEauthorrefmark{2}Department of Research \& Development, Vestel, Manisa, 45030 Turkey}
\IEEEauthorblockA{\IEEEauthorrefmark{3}Department of Electrical and Electronics Engineering, Istanbul Medipol University, Istanbul, 34810 Turkey}
\IEEEauthorblockA{\IEEEauthorrefmark{4}Department of Electrical Engineering, University of South Florida, Tampa, FL, 33620 USA}
\\
E-mails: mehmetali.aygul@ieee.org, ebubekir.memisoglu@std.medipol.edu.tr, huseyinarslan@medipol.edu.tr

\thanks{This work has been submitted to the IEEE for possible publication. Copyright may be transferred without notice, after which this version may no longer be accessible.}}

\maketitle

\begin{abstract}
Radio-frequency (RF) front-end forms a critical part of any radio system, defining its cost as well as communication performance. However, these components frequently exhibit non-ideal behavior, referred to as impairments, due to the imperfections in the manufacturing/design process. Most of the designers rely on simplified closed-form models to estimate these impairments. On the other hand, these models do not holistically or accurately capture the effects of real-world RF front-end components. Recently, machine learning-based algorithms have been proposed to estimate these impairments. However, these algorithms are not capable of estimating multiple RF impairments jointly, which leads to limited estimation accuracy. In this paper, the joint estimation of multiple RF impairments by exploiting the relationship between them is proposed. To do this, a deep multi-task learning-based algorithm is designed. Extensive simulation results reveal that the performance of the proposed joint RF impairments estimation algorithm is superior to the conventional individual estimations in terms of mean-square error. Moreover, the proposed algorithm removes the need of training multiple models for estimating the different impairments.
\end{abstract}


\section{Introduction}

\par The sixth-generation (6G) of mobile communication envisages even more diversity of applications compared to the previous generations. This will result in its application in various domains such as education, entertainment, banking, retails, healthcare, etc. \cite{de2021survey}. This necessitates new enablers including higher carrier frequencies, the accompanying larger bandwidths, higher modulation orders, massive multiple-input multiple-output (MIMO) technology, etc. However, these new enablers make the wireless communication systems more sensitive to radio-frequency (RF) impairments \cite{valkama201115}. Communication system engineers and researchers should take into consideration the effect of these impairments while designing the next generation wireless systems.

\par The RF impairments can be studied from various perspectives in wireless systems. For instance, \cite{fernandez2019machine, hilburn2018nd, sygletos2019nonlinearity} look at the impairments from a decoding aspect, while \cite{jaraut2018composite,tripathi2019characterization, paul2008physical} try to utilize the impairments for predistortion, hardware problem identification, and authentication aspect, respectively. In the case of \cite{fernandez2019machine, hilburn2018nd, sygletos2019nonlinearity}, since the target is to remove the effect of these impairments at the receiver, there is no need to estimate the individual impairments. This is because the transmitted symbols can be decoded successfully as long as the sum of multiplicative and additive impairments (joint effect of multiple impairments) are estimated and compensated. However, in the latter case where the objective is to use these impairments for predistortion, hardware problem identification, or authentication perspectives, impairments should be estimated individually to have better granularity.

\par In the literature, many algorithms use the statistical properties of the signals to estimate the RF impairments individually. Some of these model-based algorithms are explained in \cite{mohammadian2021rf}. Recent literature shows that some of these impairments have non-stationary characteristics. Hence, the aforementioned algorithms may not always be capable of addressing this issue. This effect becomes more pronounced with the aforementioned enablers, such as higher modulation orders and frequencies, envisaged in 6G and beyond communication systems \cite{giordani2020toward}. In view of these challenges, machine learning (ML) and deep learning (DL)-based algorithms have been proposed for more sophisticated RF impairment estimation.

\begin{figure*}[!ht]
\centering
\resizebox{1.95\columnwidth}{!}{
\includegraphics[width=14cm]{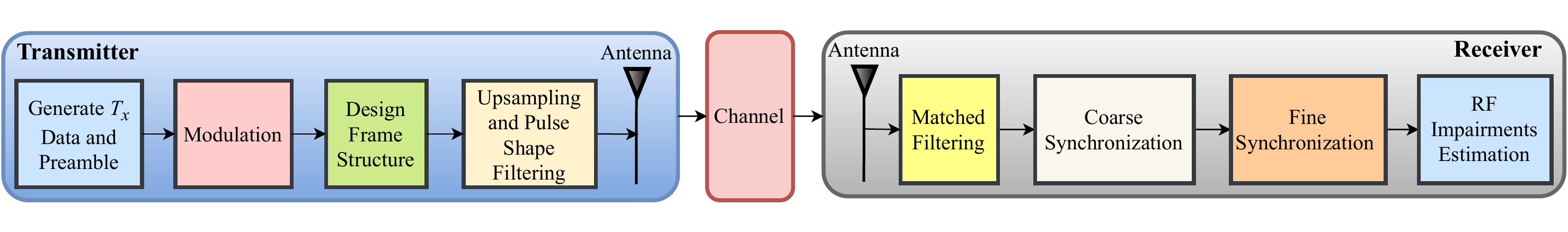}}
\vspace{-0.5cm}
\caption{A block diagram of data transmission in a wireless communication system.}
\label{system_model}
\end{figure*}

\par Several algorithms are developed to estimate RF impairments with ML-based algorithms \cite{zibar2015application, wong2018estimation, wong2019specific, patent1, zhou2019coarse, dreifuerst2020deep}. Bayesian filtering is used to estimate phase noise \cite{zibar2015application}. Convolutional neural networks are used to estimate the transmitter in-phase ($I$) and quadrature ($Q$) path gain imbalance using raw $I/Q$ data as an input \cite{wong2018estimation,wong2019specific}. A reinforcement learning-based algorithm is proposed to estimate the same in \cite{patent1}. Frequency offset estimation is addressed in \cite{zhou2019coarse, dreifuerst2020deep} where the former provides a neural-network-based solution for MIMO systems, while the latter proposes a DL algorithm for estimating the same in millimeter-wave and terahertz frequencies systems. However, these algorithms estimate the impairments individually and do not utilize the existing relationship between the impairments, which may limit the estimation accuracy if impairments have similar effects on the received signals.

\par The joint estimation of multiple RF impairments is especially suitable for cases where the different impairments have a joint (additive and/or multiplicative) effect on the signal. However, joint estimation is a complex problem, and it is difficult to have an accurate estimation performance. It is intuitively sound to think of using a DL algorithm to solve complex problems \cite{najafabadi2015deep}. Especially, DL-based multi-task learning algorithms \cite{caruana1997multitask} are capable of addressing complex joint estimation problems since they can utilize the relationship between related tasks through their hidden layers. Along with this line, different from the existing literature, this paper proposes to estimate multiple RF impairments jointly for improving their estimation accuracy by using relationships between different impairments. For this purpose, a DL-based multi-task learning algorithm is designed. Extensive simulation results show a performance improvement achieved by the proposed algorithm compared to DL-based single RF impairment estimation in terms of mean square error (MSE). Furthermore, instead of training a DL model for each impairment, a single DL model is used in the proposed algorithm.

\section{System Model and Preliminaries}
\label{Section2}
\subsection{System Model}

\par
\begin{figure}[!htp]
\centering
\resizebox{.7\columnwidth}{!}{
\includegraphics[width=14cm]{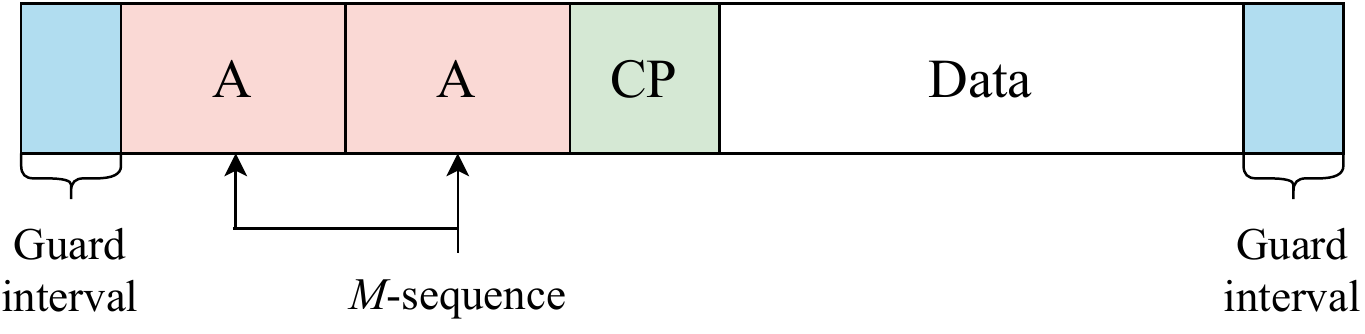}}
\vspace{-0.3cm}
\caption{Illustration of the designed frame.}
\label{framewaveform}
\end{figure}

 In this paper, single-carrier modulation is considered as in \cite{pancaldi2008single}. The system model can be divided into three blocks; transmitter, channel, and receiver. These blocks are illustrated in Fig.~\ref{system_model}. In the transmitter, transmitted data $T_x$ and the preambles with $M$-sequence length are generated. Then, the modulated symbols are designed as the frame structure in Fig.~\ref{framewaveform}. Here note that the cyclic prefix (CP) is used to avoid inter-symbol interference and provide circular convolution. Afterward, the signal is upsampled and filtered, and send through the channel. In the channel, the frequency selective channel with additive white Gaussian noise (AWGN) distorts the transmitted signal. In the receiver, the matched filter is used to maximize the signal-to-noise ratio (SNR) and the coarse synchronization is performed to find the starting point of the frame. Then, fine synchronization is performed and RF impairments are estimated.

\vspace{-0.2cm}

\subsection{The RF Impairments}

\par Communication systems suffer from a wide variety of impairments. Some of these impairments (their parameters); $I/Q$ gain imbalance ($I_g$/$Q_g$), quadrature offset ($\psi$), phase noise ($\phi$), and $I/Q$ offset ($I_o$/$Q_o$); are estimated in this paper. These impairments are briefly explained below, and their effects are illustrated in Fig.~\ref{hardware_impairments} with their components \cite{arslan, arslan2}. In this figure, parameters of the RF impairments are selected as follows; $I_g=1.3$, $Q_g=0.8$, $\psi=0.9$, $\phi=\pi/3$, $I_o=0.21$, and $Q_o=-0.15$.

\subsubsection{$I/Q$ gain imbalance} It causes the in-phase component of the signal $(I)$ to be smaller than the quadrature component of the signal $(Q)$, or vice versa. This results in an imbalanced change in the original constellation. The imbalance of $I/Q$ gain can be measured as $(I_g/Q_g -1)\times 100$ in percent.

\subsubsection{Quadrature offset} It causes due to the phase offset in the sine and cosine signal generation. This distorts the orthogonality between $I/Q$ branches so that the degree between the branches is not 90$^{\circ}$. The signal with quadrature offset is modeled as  
\begin{equation}
x_{qo} (t) = \cos{(2\pi f_ct)}I_g\Re\{x(n)\}+ \sin{(2\pi f_ct+\psi)}Q_g\Im\{x(n)\},
\end{equation}
\noindent where $f_c$ and $x(t)$ denote the carrier frequency and modulated time-domain signal after DAC and band-bass filter processes, respectively. $\Re$ and $\Im$ represent real and imaginary part of the data, respectively. Also, $t$ and $n$ represent the time and sample, respectively.

\subsubsection{Phase noise} It is caused by the spectrum spread at the desired spectrum of a practical local oscillator. It is assumed that the phase noise is constant over the data so that the effect of the phase noise is only the phase rotation of the symbols. This can be described as 
\begin{equation}
x_{pn} (t) = x_{qo} (t)e^{j\phi},
\end{equation}
\noindent where $\phi$ denotes the phase rotation.

\subsubsection{$I/Q$ offset} It is essentially caused by the mixers and direct current signals. The effect is a shift in the constellation. The signal with $I/Q$ offset is modeled as
\begin{equation}
x_{iqo} (t) = x_{pn} (t) + I_{o} + jQ_{o},
\end{equation}
\noindent where $I_{o}$ and $Q_{o}$ denote the offsets for the $I$ and $Q$ components.

\par
\begin{figure}[!t]
\centering
\resizebox{.99\columnwidth}{!}{
\includegraphics[width=14cm]{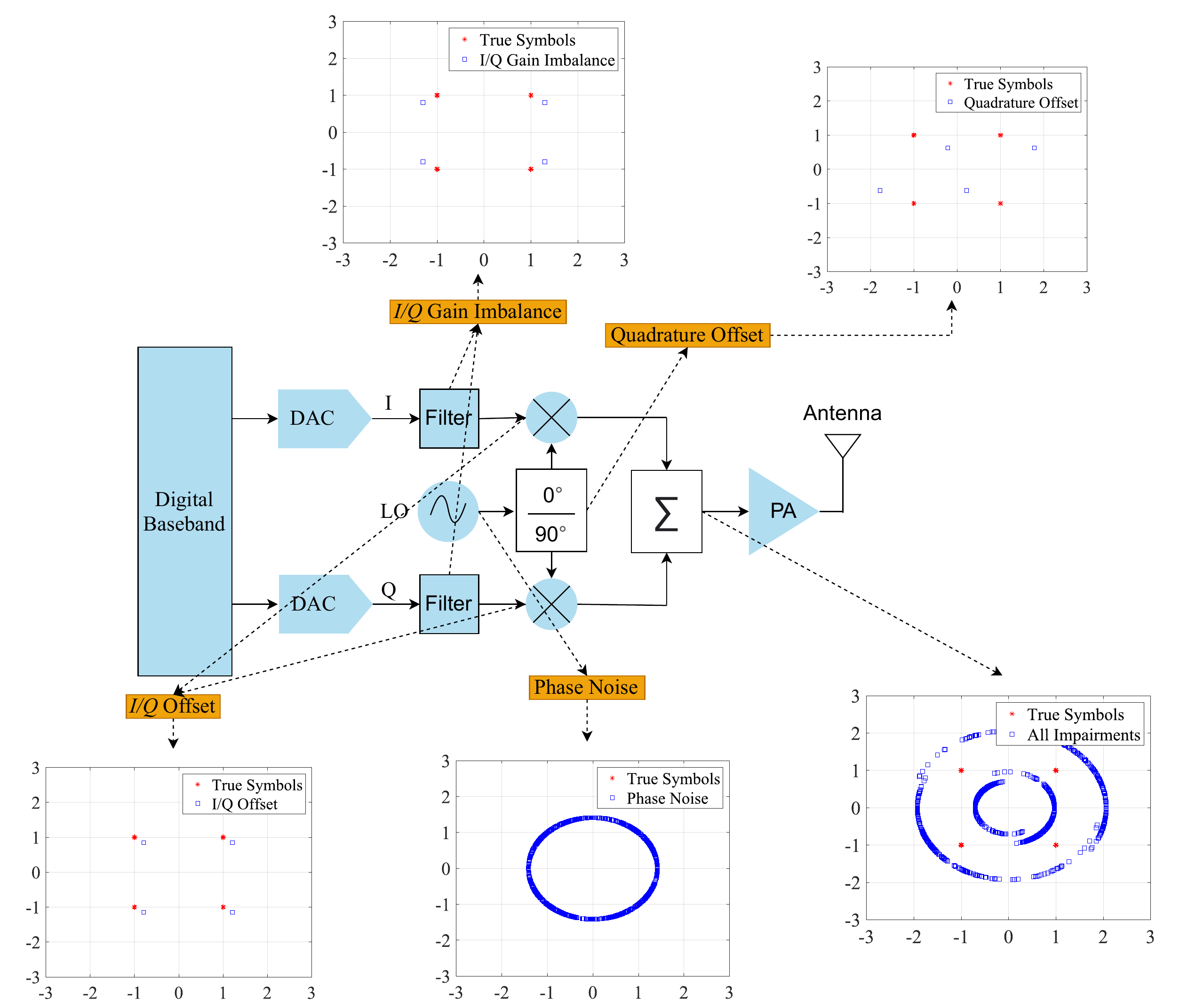}}
\caption{Typical transceiver chain with various sources of RF impairments.}
\label{hardware_impairments}
\end{figure}

 When all of the aforementioned impairments are added to the transmitted signal, the modulated symbols are seen as the constellation plot in Fig.~\ref{hardware_impairments}. As seen from this figure, when the effect of impairments is mixed, the constellation diagram becomes more distorted, which makes the estimation problem complex and difficult.

\par After the signal with all these impairments passes throughout the wireless channel, the baseband received time-domain signal can be defined as 
\begin{equation}
    r(n) = \kappa(n)x(n)+ \zeta(n) + w(n),
\end{equation}
\noindent where $\kappa(n)$, $\zeta(n)$, and $w(n)$ denote the multiplicative impairments, the additive impairments, respectively and the complex Gaussian sample with the distribution of $\mathcal{CN}(0,\sigma_{N}^{2})$. Here, the channel is considered as frequency selective, and the maximum excess delay is smaller than the CP duration.

\subsection{Deep Learning}

\par There is a wide variety of successful ML algorithms applications in diverse areas ranging from image and speech processing to pattern recognition. This success motivated their applicability to the area of wireless communication \cite{jiang2016machine}. These algorithms are expected to become an indispensable part of 6G and beyond wireless communication systems.

\par Amongst ML algorithms, DL-based algorithms have become popular since the usage of multiple hidden layers of DL algorithms enables to magnify the intrinsic distinctive data features while suppressing the irrelevant information at each layer \cite{lecun1990handwritten}. This is especially the case for complex problems where multiple problems are tried to estimate jointly and the system becomes blind \cite{ali2021blind}. Therefore, these algorithms are suitable candidates for complex problems. Also, in these algorithms, raw data can be used without specific feature engineering/crafting thanks to the aforementioned added benefits of the deep architectures.

\section{Joint RF Impairments Estimation}
\label{Section3}

\subsection{The Proposed Algorithm}

\par The RF front-end components cause multiplicative and additive impairments in the communication systems, as described in (4). In this model, if only one RF impairment is considered in the system, the estimation with the classical algorithms can be performed optimally. However, multiple RF impairments are considered as a practical scenario in the system model, as shown in Fig.~\ref{hardware_impairments}. Therefore, the accurate estimation of the impairments caused by the individual components becomes a challenging problem with classical estimation algorithms. For instance, let's assume that the phase noise and $I/Q$ offset exist in (4), then the received signal considering the AWGN channel can be written as
\begin{equation}
    r(n) = x(n)e^{j\phi} + I_{o} + jQ_{o} + w(n).
\end{equation}
\noindent This equation shows that when the estimation of the phase offset is made from $r(n)$, the accuracy of the estimation does not only depend on the noise but both noise and $I/Q$ impairment. Moreover, when the number of RF components causing an impairment increases, the problem becomes kind of a black box, and accurate estimation of individual impairment becomes unfeasible with the classical algorithms.

\par DL-based algorithms are highly preferred in black-box problems. Still, in DL, optimization is made based on a certain metric, whether this is a score on a particular benchmark or a business key performance indicator. For this optimization, a single model or an ensemble of models are generally trained to perform the desired task. While an acceptable performance is generally achieved in this way, by being laser-focused on a single task, trying to find optimum solutions simultaneously for different tasks may increase the performance of the original task. In other words, our model can estimate the original problem better, by sharing representations between related tasks.

\par Multi-task learning is well-suited to estimate multiple impairments since it can leverage useful information contained in multiple related tasks to improve the individual estimates \cite{zhang2017survey}. This is especially the case where it is used with several hidden layers (DL-based multi-task learning algorithm). In view of these discussions, the usage of a DL-based multi-task learning algorithm is proposed in this paper to estimate multiple RF impairments jointly.

\begin{figure}[!t]
\centering
\begin{tabular}{@{}c@{}}
\includegraphics[width=.9\linewidth]{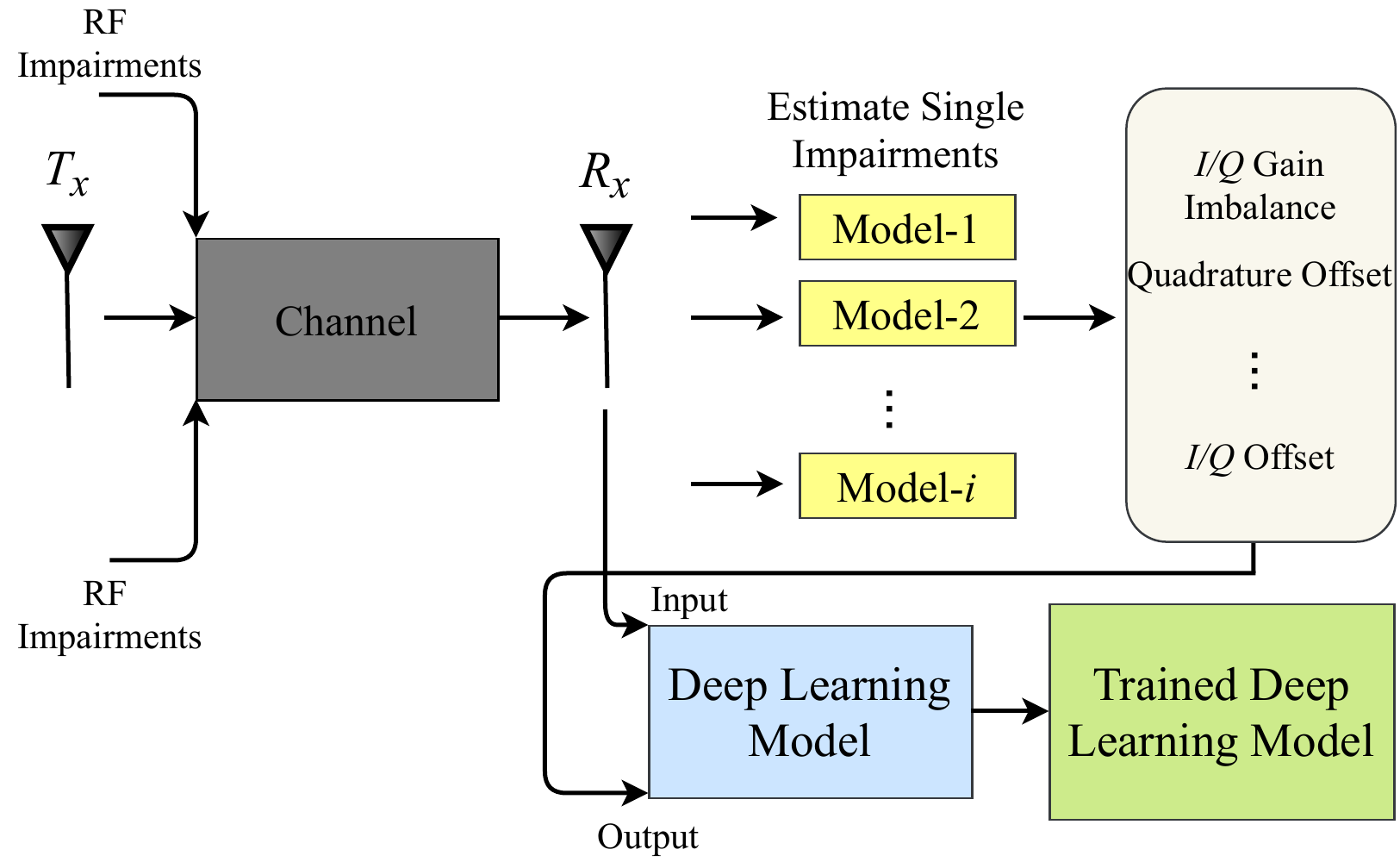} \\
\small (a) 
\end{tabular}
\begin{tabular}{c}
\includegraphics[width=.95\linewidth]{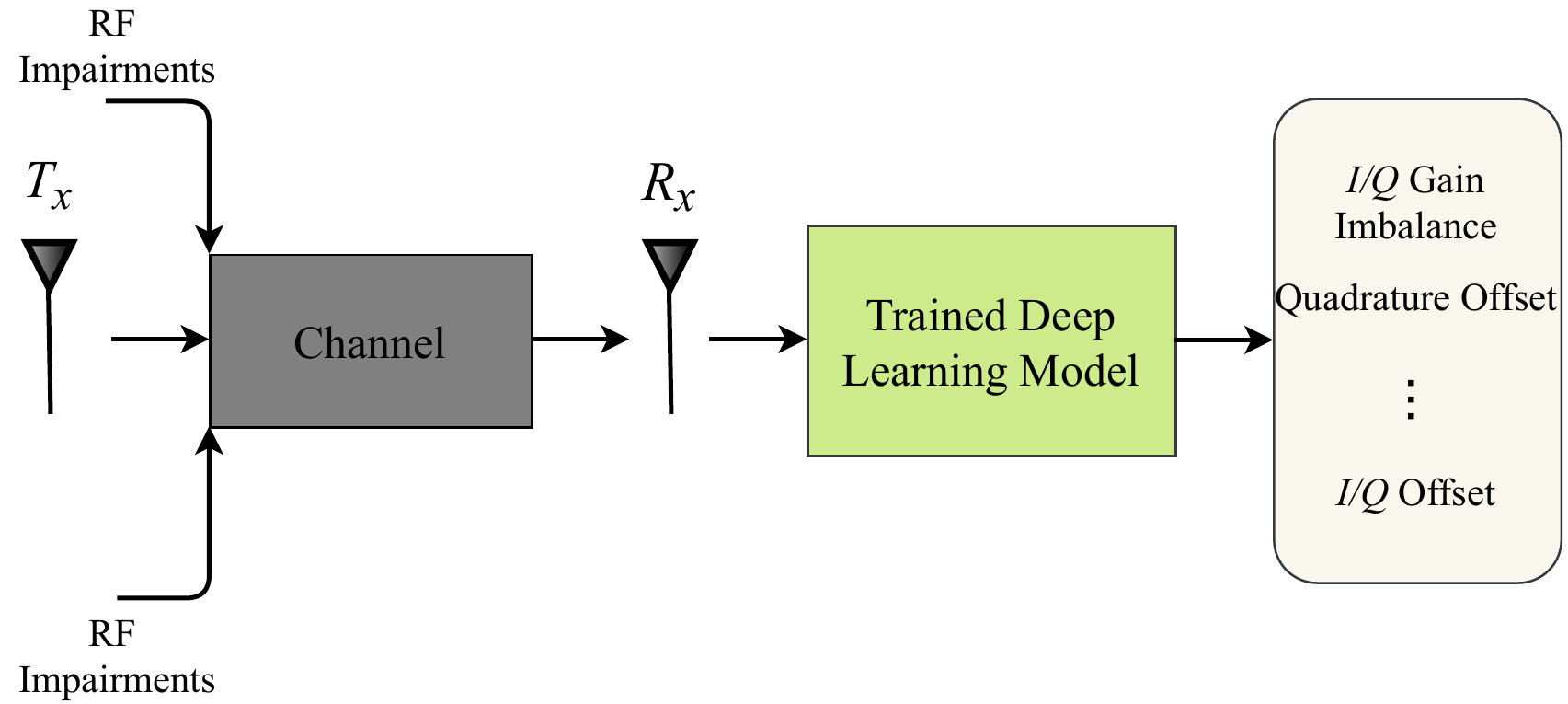} \\
\small (b) 
\end{tabular}
\vspace{-0.2cm}
\caption{The proposed algorithm for RF impairments estimation; (a) training and (b) testing stages.}
\label{proposed_algorithm}
\end{figure}

\par The proposed algorithm based on DL-based multi-task learning operates in two stages. First, the proposed algorithm is performed in the training stage, where the dataset is generated, and a DL algorithm is configured and trained. Afterward, it is performed in the testing stage where multiple RF impairments are estimated simultaneously.

\begin{figure}[!b]
\centering
\centering\resizebox{0.87\columnwidth}{!}{
\includegraphics{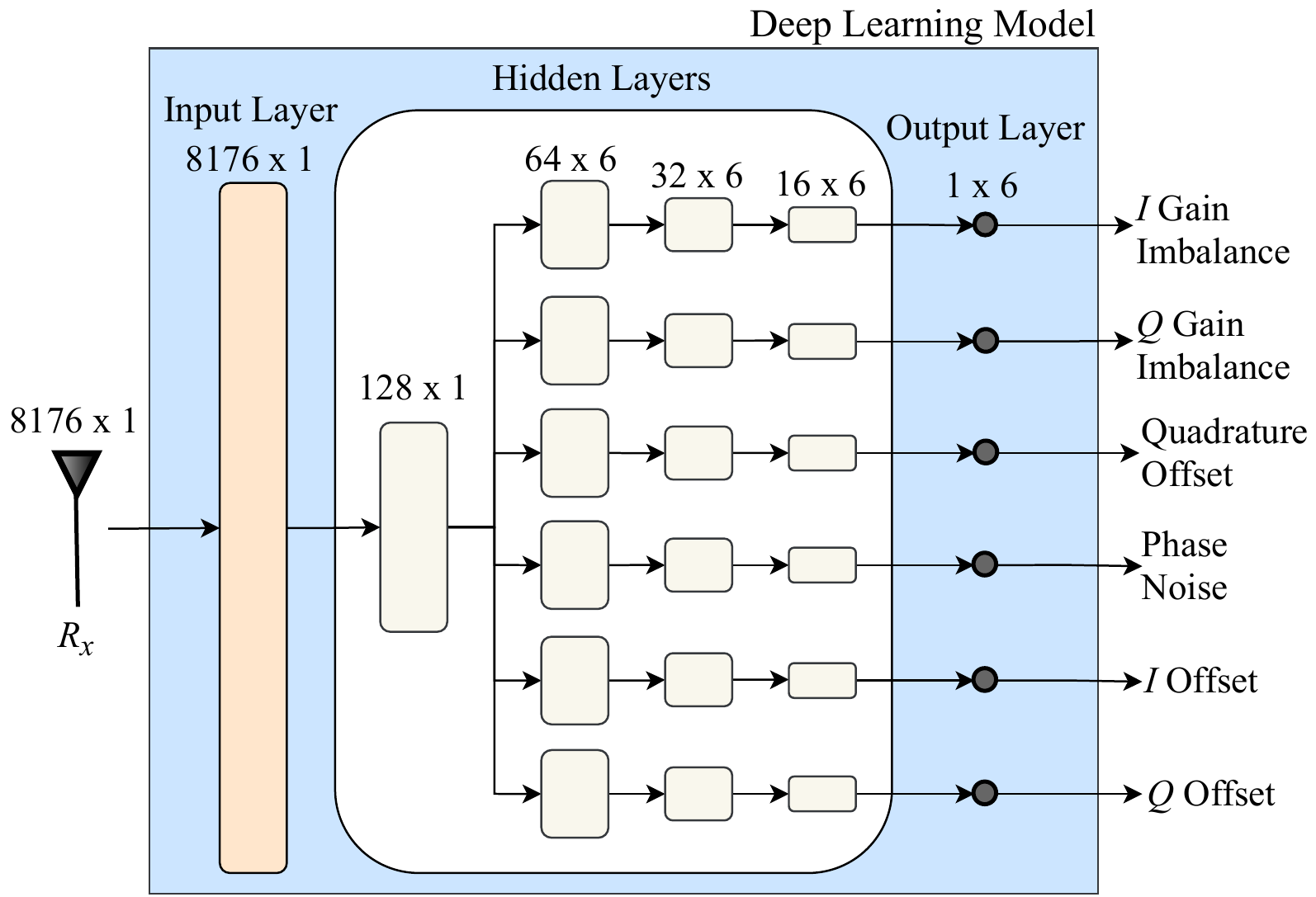}}
\vspace{-0.2cm}
\caption{A DL-based multi-task learning algorithm for joint RF impairments estimation.}
\label{mlagorithm}
\end{figure}

\par In the training stage, $T_x$ signals are transmitted via a wireless communication channel. Then, $R_x$ is captured at the receiver based on the system model explained in Section~\ref{Section2}. Afterward, various impairments, such as $I/Q$ gain imbalance, quadrature offset, phase noise, and $I/Q$ offset or their parameters are estimated by conventional algorithms (model or ML-based algorithms). Here note that the training dataset can be also obtained by a computer or test equipment. Then, these estimated values are stored as output in vector format. Correspondingly, the received signals, where these values are obtained, are stored as the input. These processes are repeated until enough amount of dataset is generated. The size of the dataset is determined according to the system requirements as it is optimum in terms of system performance, complexity, and memory. Then, the DL algorithm\footnote{All hyperparameters of the DL algorithm are tuned empirically by considering the performance and generalization capability of the proposed algorithm.} is trained with the created dataset. These processes are illustrated in Fig.~\ref{proposed_algorithm} (a). Also, an example of the DL-based multi task learning algorithm is illustrated in Fig.~\ref{mlagorithm}. Once the training and validation\footnote{The validation dataset is generally used in the ML context to provide an unbiased evaluation of a model fit on the training dataset \cite{ethemalpaydin}.} loss convergence is done in the training stage, the testing stage starts, which characterizes the run-time operation of the algorithm.

\par In the testing stage, a signal, which is distorted by the wireless channel and RF impairments, is captured in the receiver. Then, this distorted signal is fed to the trained DL algorithm. Afterward, the trained DL algorithm estimates the multiple RF impairments. These processes are illustrated in Fig.~\ref{proposed_algorithm} (b). Overall processes of the proposed algorithm can be found in Algorithm 1.

\begin{figure}[!t]
\vspace{0.0001pt}
\end{figure}

\begin{algorithm}[!t]
\caption{Estimating multiple impairments with a single model.}
\label{Algorithm1}
\begin{algorithmic}[1]{
\renewcommand{\algorithmicrequire}{\textbf{Input:}}
\renewcommand{\algorithmicensure}{\textbf{Output:}}
\REQUIRE $S$ number of training received signals ($R_{xtrain}$), $N$ number of validating received signals ($R_{xvalidation}$), initial hyperparameters, conventional algorithms to estimate single-RF impairments, and testing received signals ($R_{xtest}$).  
\ENSURE Estimated RF impairments ($E_{test}$). \\
\textbf{Training Stage:}\\
\FOR{$s = 1$ \textbf{to} $S$}
 \STATE{Receive $R_{xtrain}$.
 }
 \STATE{Conventional algorithms estimate each RF impairments ($E_{train}$).
 }
 \STATE{A new data point $R_{xtrain}$ and $E_{train}$ is added to the training dataset ($D$).
 }
 \ENDFOR
}
    \STATE{Train the DL algorithm using the generated dataset $D$.}    
   \WHILE{the training and validation loss graphs are converged}{
    \STATE{Change the hyperparameters of the DL algorithm.}
    \STATE{Train the DL algorithm using the generated dataset $D$.}    
}
  \ENDWHILE
  \\
\textbf{Testing Stage:}\\ 
 \STATE{Receive $R_{xtest}$.
 }
 \STATE{Estimate multiple RF impairments $E_{test}$ using $R_{xtest}$ and trained DL algorithm.
 }
\end{algorithmic}
\end{algorithm}

\subsection{A Note on Computational Complexity}

\par Training and testing stages determine the computational complexity of the proposed algorithm. The complexity of the training stage is based on both the model-based estimations and the DL algorithm, while the testing stage complexity solely depends on the DL algorithm.

\par A DL-based multi-task learning algorithm is used in this work with an input layer, four hidden layers, and an output layer. This algorithm has $a$ units in the input layer, where $a$ represents the size of the input vector. Also, it has $b$ hidden units for joint learning. Besides that it has $c$, $d$, and $e$ hidden units and $f$ output units for each impairment. Therefore, the overall training computational complexity of this algorithm is $\bigO(ml\times (ab+6(bc)+6(cd)+6(de)+6(ef)))$, where $m$ and $l$ represents number of epochs and training examples, respectively. Also, the computational complexity of the validation and the number of trials to select optimum hyperparameters of the DL algorithm can be added to the training complexity. Note that, number of trials and validation data depends on the complexity and reliability requirements of the application. The per-sample computational complexity of testing is around half that of the training stage since the testing stage does not need back-propagation \cite{he2015convolutional}.

\section{Simulation Results}
\label{Section4}

\subsection{Parameter Settings}

\par Dataset\footnote{Dataset will be publicly available online after the acceptance.} is generated by MATLAB simulation environment depending on the setup explained in Section \ref{Section2}. In this setup, all transmitter parameters such as modulation type, roll of factor of root-raised-cosine (RRC) filter, data length, etc., are assumed to be known by the receiver. Detailed specifications of the simulation parameters are listed in Table~\ref{Simulation_parameters}. In these simulations, only $I/Q$ gain imbalance, quadrature offset, phase noise, and $I/Q$ offset impairments are estimated. The values of these parameters and other imperfections added to the system are given in Table~\ref{Impairment_parameters}\footnote{Impairments are randomly selected between the given values for each sample.}. Here note that it is possible to estimate any kind of impairment in the system, without changing the architecture, by training the network according to newly added impairment.

\par The dataset is split into three sets; training, validation, and testing. In the training stage, 30000 samples are used. Here note that, in the training stage, SNR is varied from 0 to 20 with a step size of 5. In other words, 6000 samples are used for each SNR value. Also, it is assumed that the true values of the impairments are known by the receiver in the training stage for the sake of simplicity. In the validation and testing stages, 10000 samples are used for each SNR value.

\par The conventional and the proposed DL algorithms for RF impairments estimation are implemented by Keras \cite{keras}, an open-source ML library under the Python environment. All of the algorithms were trained and tested on an MSI computer with an Intel\textregistered\ Core\texttrademark\ i7-7700HQ central processing unit (CPU) \text{@} 2.80 GHz CPU, GeForce GTX 1050 Ti graphics processing unit (GPU), 16 GB RAM, and Windows 10 operating system.

\vspace{-0.2cm}
\subsection{Hyperparameters of the Deep Learning Algorithms}

\par An input layer, four hidden layers (fully connected layers), and an output layer are used in the proposed algorithm. Particularly, 8176 units are used in the input layer. Then, joint learning of the impairments is made in the first hidden layer with 128 units to learn the relationship between the impairments. Afterward, three hidden layers are used to learn characteristics of each impairments. In these layers, 64, 32, and 16 units are used, respectively. Lastly, all of the impairments are estimated in the output layer with a unit per impairment. In all of these layers, the rectified linear unit is used as an activation function. In total, 1090214 parameters are used in the proposed DL algorithm and the DL algorithm is trained with a batch size of 16 and 40 epochs. ADAM \cite{kingma2014adam} is used for adaptive learning rate optimization and the optimum learning rate in this algorithm was found at 0.00001.

\begin{table}[t]
\centering
\caption{Simulation parameters.}
\vspace{-0.2cm}
\resizebox{0.86\columnwidth}{!}{
\begin{tabular}{|c|c|}
\hline
\textbf{Parameter}                                 & \textbf{Value} \\ \hline
Number of symbol blocks                            & 10             \\ \hline
Number of symbols in each block                    & 64            \\ \hline
Cyclic prefix ratio compared to block size         & 1/4            \\ \hline
Oversampling rate                                 & 4             \\ \hline
Sampling frequency (sample/second)                                        & $4\times 10^6$           
\\ \hline
$M$-sequence length                                        & $2^{9}-1$      
\\ \hline
Modulation order                                   & QPSK           \\ \hline
Guard interval                                & 245            \\ \hline
Filter type                                        & RRC            \\ \hline
Roll of Factor                                     & 0.3            \\ \hline
\end{tabular}}
\label{Simulation_parameters}
\end{table}

\begin{table}[!t]
\centering
\caption{RF impairments parameters.}
\vspace{-0.2cm}
\resizebox{0.68\columnwidth}{!}{
\begin{tabular}{|c|c|}
\hline
\textbf{Impairment}   & \textbf{Range}       \\ \hline
Channel model         & Rayleigh channel     \\ \hline
Noise                 & AWGN                 \\ \hline
$I/Q$ gain imbalance  & 0 to 1.5             \\ \hline
Quadrature offset     & -1 to 1              \\ \hline
Phase noise           & 0 to $\pi/2$         \\ \hline
$I/Q$ offset          & -0.5 to 0.5          \\ \hline
\end{tabular}}
\label{Impairment_parameters}
\end{table}

\par For the conventional algorithm, six different DL algorithms are implemented. In these algorithms, an input layer, four hidden layers, and an output layer are used. In the input layer 8176 units, in the hidden layers, 128, 64, 32, and 16 hidden units are used, while a unit is used in the output layer. In total, 1057537 parameters are used for each impairment estimation (each DL algorithm). Other all hyperparameters are the same as the proposed algorithm.

\begin{figure}[!b]
\centering
\resizebox{0.99\columnwidth}{!}{
\begin{tabular}{cc}
\includegraphics{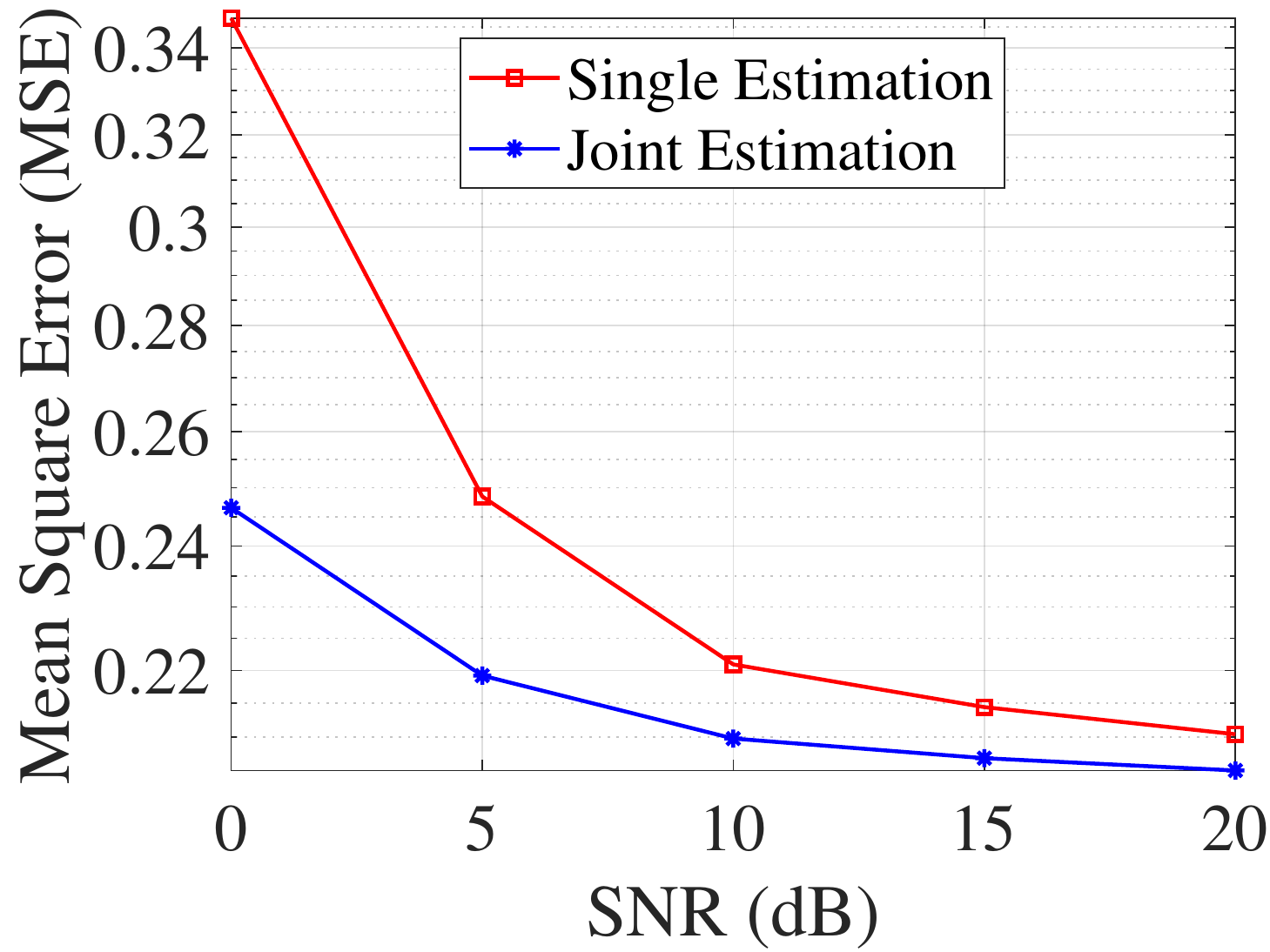}&
\includegraphics{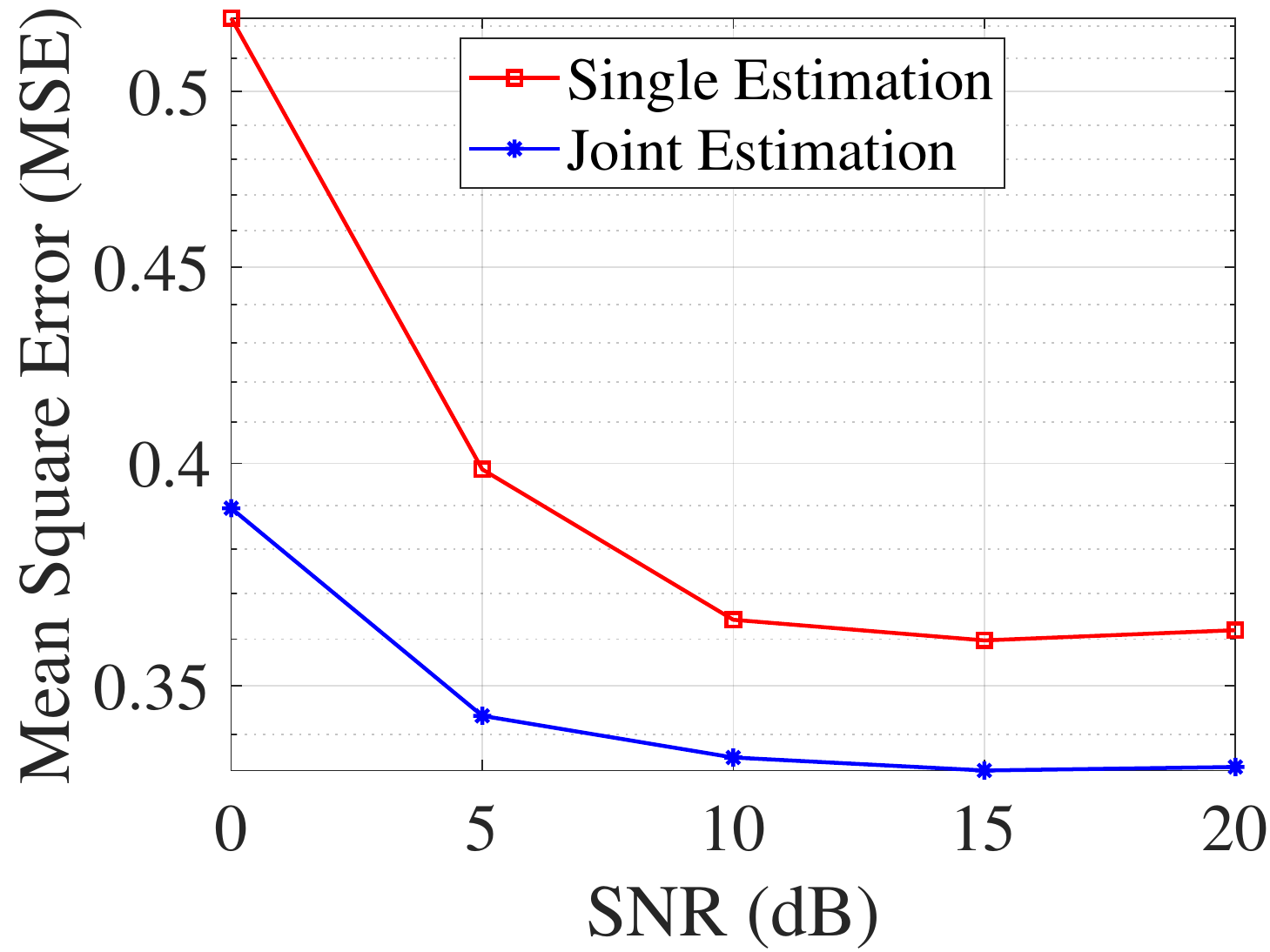} \\
\hspace{+2cm}\Huge{(a)} & \hspace{+2cm}\Huge{(b)}  \\
\includegraphics{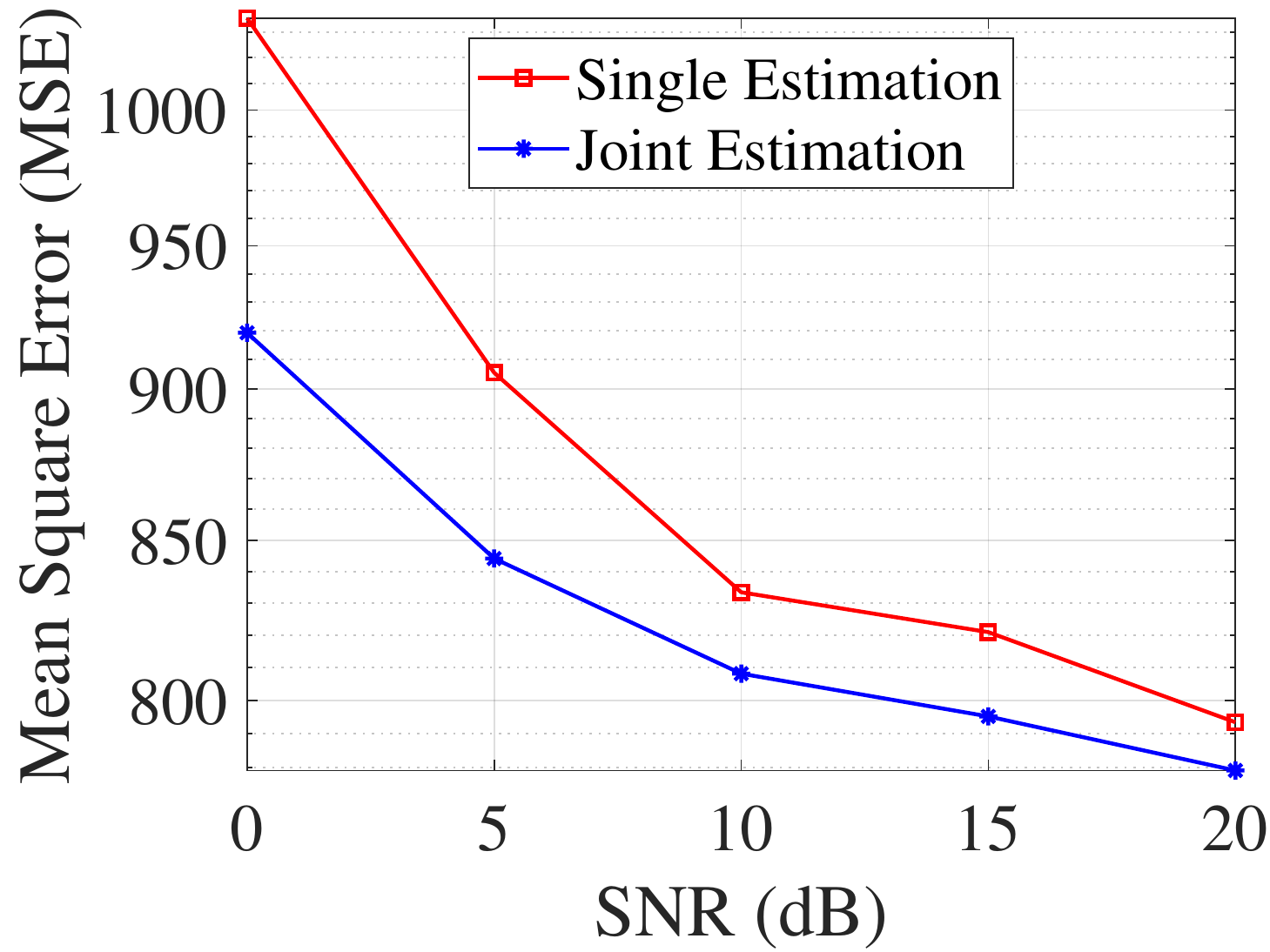}&
\includegraphics{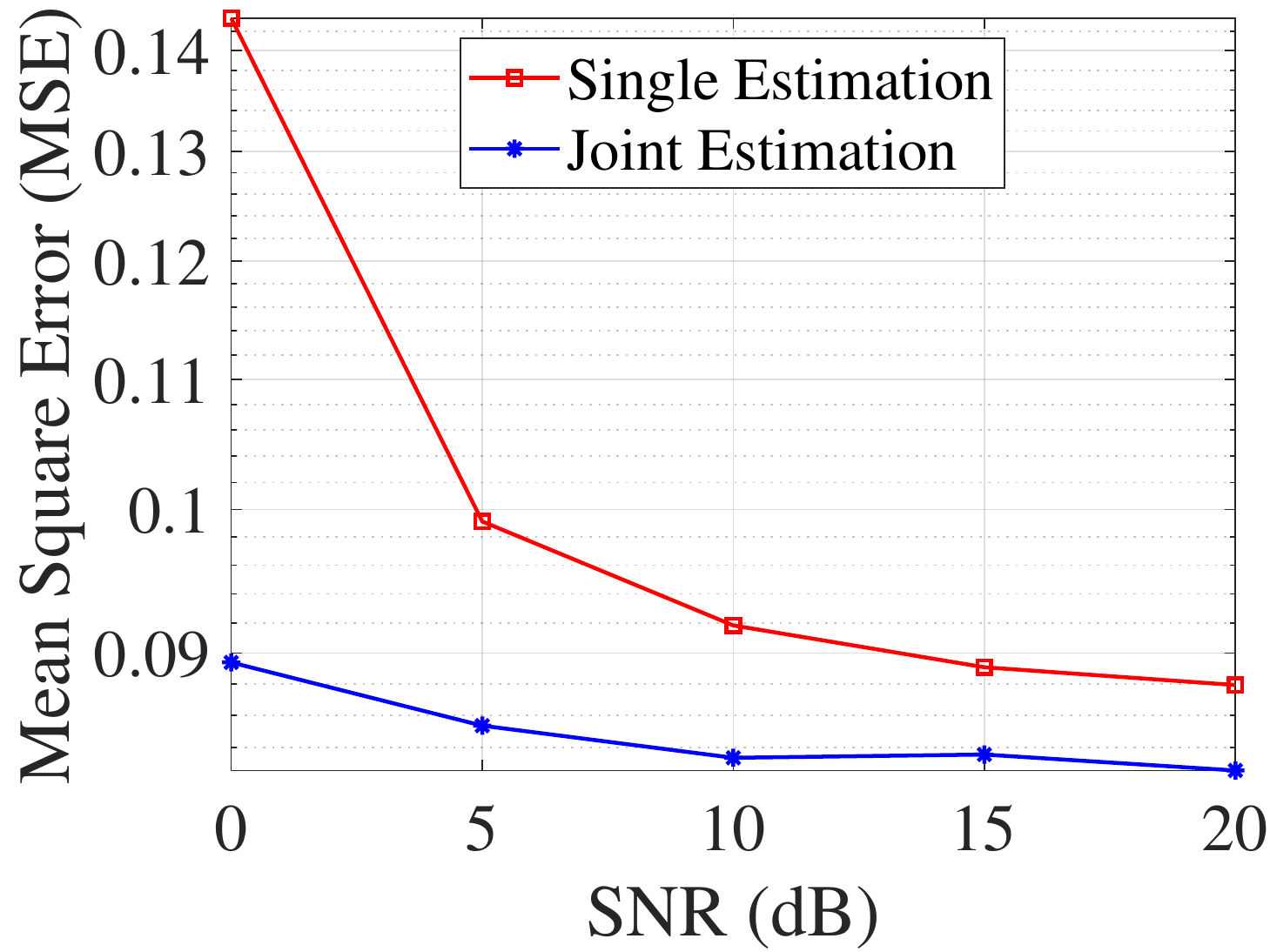}\\
\hspace{+2cm}\Huge{(c)} & \hspace{+2cm}\Huge{(d)} 
\end{tabular}}
\caption{MSE results for (a) $I/Q$ gain imbalance estimation, (b) quadrature offset estimation, (c) phase noise estimation, and (d) $I/Q$ offset estimation.}
\label{mse_all} 
\end{figure}

\subsection{Performance Evaluation}

\par The performance of the conventional algorithm ($Single\:Estimation$), where each DL algorithm estimates one impairment at a time, and the proposed algorithm ($Joint\:Estimation$), where the DL algorithm estimates all impairments jointly, is compared in Fig.~\ref{mse_all}. These analyses are made by comparing the MSE of the estimations. Here note that for the sake of simplicity, estimation of the corresponding parameters of each impairment is compared to evaluate the performance of the algorithms. Fig.~\ref{mse_all} demonstrates that the performance of the proposed algorithm is superior to the conventional algorithm. This is consistently true for all impairments ($I/Q$ gain imbalance, $I/Q$ offset, quadrature offset, and phase noise) and SNR values.

\par
\begin{figure}[t!] 
\centering\resizebox{0.65\columnwidth}{!}{
\includegraphics{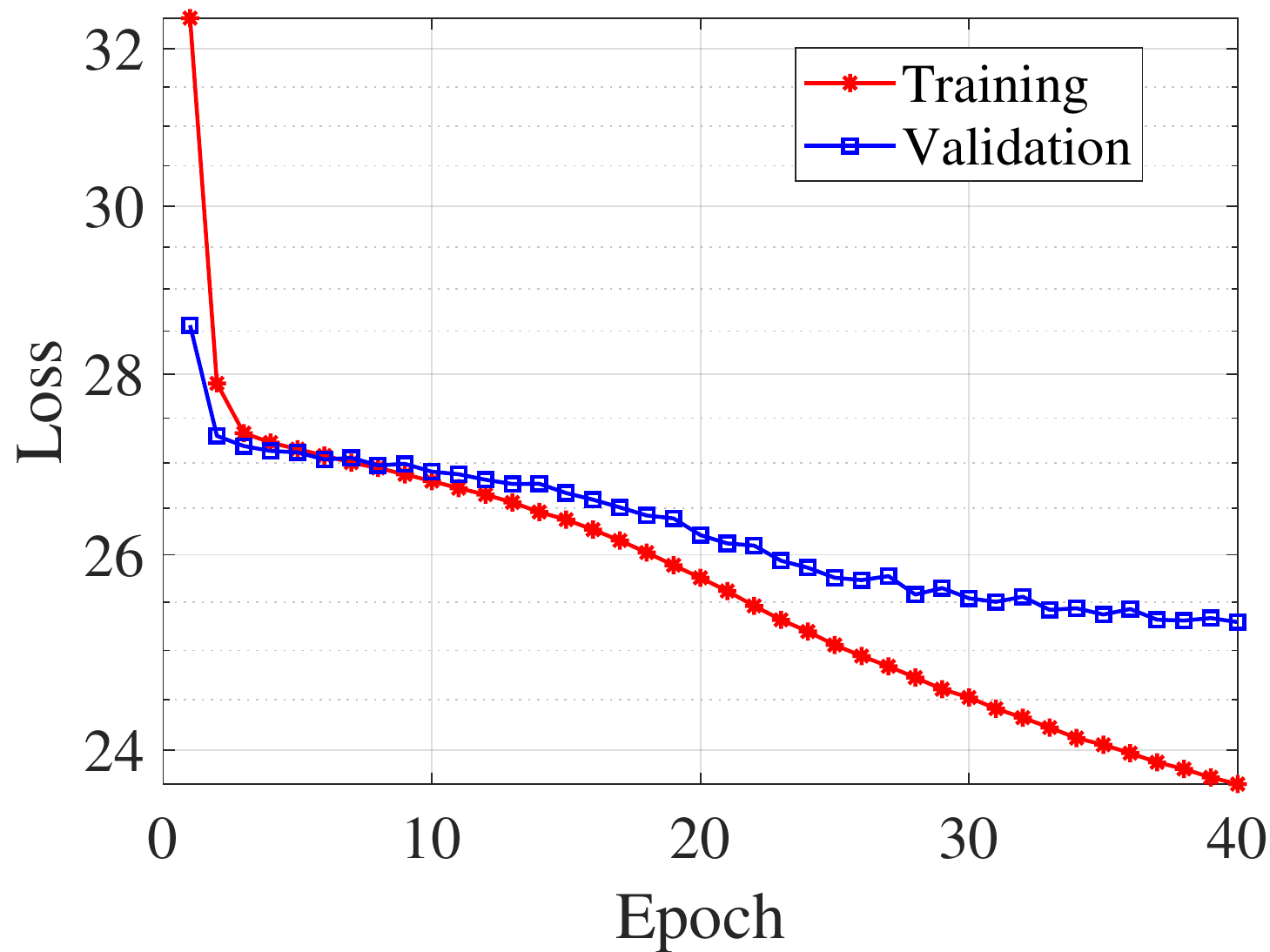}}
\vspace{-0.3cm}
\caption{The loss graph of the DL algorithm.}
\label{loss}
\end{figure}

 As the proposed algorithm is based on DL, it is crucial to ensure that the developed algorithm is generalized well \cite{Goodfellow}, i.e., the inputs are not memorized during the training stage. In order to observe this, the training and validation losses versus epochs for joint RF impairments estimation problem is presented in Fig.~\ref{loss}. This figure demonstrates that the training sets converges to the validation set. This means there is no overfitting during the training which demonstrates the ability to work with unforeseen data for the proposed algorithm.

\section{Conclusions}
\label{Section5}

\par This paper proposed the joint estimation of multiple RF impairments to exploit the relationship between them. For this purpose, a DL-based multi-task learning algorithm was designed. This algorithm estimated the RF impairments of $I/Q$ gain imbalance, $I/Q$ offset, quadrature offset, and phase noise with a single model. Thus, there was no need to train multiple models to estimate different impairments. Simulation results were shown that estimating multiple-RF impairments jointly improves the estimation accuracy, as tested over a realistic scenario. As future work, the DL-based joint RF impairments estimation will be investigated from the perspectives of predistortion, hardware problem identification, and physical layer authentication. Also, the performance of the proposed algorithm will be evaluated in the real-world environment.

\section*{Acknowledgment}
\par This work was supported in part by the Scientific and Technological Research Council of Turkey (TÜBİTAK) under Grant No. 5200030 with the cooperation of VESTEL and Istanbul Medipol University. Also, the authors would like to thank Muhammad Sohaib J. Solaija for his valuable comments and suggestions to improve the quality of the paper.

\bibliographystyle{IEEEtran}

\end{document}